\documentclass[iop, numberedappendix]{emulateapj}

\usepackage[colorlinks,urlcolor=blue,citecolor=black,linkcolor=blue]{hyperref}
\usepackage{amsmath}
\usepackage[export]{adjustbox}

\newcommand{\h}{\ensuremath{{\rm H}}}
\newcommand{\hi}{H\,{\sc{i}}}

\newcommand{\oi}{O\,{\sc{i}}}
\newcommand{\water}{\ensuremath{{\rm H_2O}}}
\newcommand{\hm}{\ensuremath{{\rm H}_2}}
\newcommand{\nm}{\ensuremath{{\rm N}_2}}
\newcommand{\oh}{\ensuremath{{\rm OH}}}

\newcommand{\otwo}{\ensuremath{{\rm O}_2}}

\newcommand{\xh}{\ensuremath{x({\rm H})}}

\newcommand{\Nh}{\ensuremath{N_{\rm H}}}

\newcommand{\xhm}{\ensuremath{x({\rm H}_2)}}
\newcommand{\xoh}{\ensuremath{x({\rm{OH}})}}

\newcommand{\Tg}{\ensuremath{T_{\rm g}}}
\newcommand{\ag}{\ensuremath{a_{\rm g}}} 	

\newcommand{\Sd}{\ensuremath{S_{\rm d}}}
\newcommand{\rhod}{\ensuremath{\rho_{\rm d}}}
\newcommand{\rhog}{\ensuremath{\rho_{\rm g}}}

\newcommand{\Lya}{\ensuremath{{\rm Ly}\alpha}}

\newcommand{\LFUV}{\ensuremath{L_{\mathrm{FUV}}}}

\newcommand{\Fcnt}{\ensuremath{F_\mathrm{cont}}}

\newcommand{\GLya}{\ensuremath{G_\mathrm{Ly\alpha}}}
\newcommand{\Gcnt}{\ensuremath{G_\mathrm{cont}}}

\newcommand{\Lx}{\ensuremath{L_{\rm X}}}
\newcommand{\Tx}{\ensuremath{T_{\rm X}}}

\newcommand{\ah}{\ensuremath{\alpha_{\rm h}}}

\newcommand{\um}{\ensuremath{\,\mu}m}
\newcommand{\Lsun}{\ensuremath{\,L_{\odot}}}
\newcommand{\LSun}{\ensuremath{\,L_{\odot}}}
\newcommand{\Msun}{\ensuremath{\,M_{\odot}}}
\newcommand{\MSun}{\ensuremath{\,M_{\odot}}}

\newcommand{\RSun}{\ensuremath{\,R_{\odot}}}
\newcommand{\erg}{\ensuremath{{\rm erg}}}
\newcommand{\ergps}{\ensuremath{{\rm erg}\,{\rm s}^{-1}}}

\newcommand{\cm}{\ensuremath{{\rm \,cm}}}
\newcommand{\sqcm}{\ensuremath{{\rm \,cm}^2}}
\newcommand{\psqcm}{\ensuremath{{\rm \,cm}^{-2}}}
\newcommand{\ps}{\ensuremath{{\rm \,s}^{-1}}}

\newcommand{\pcc}{\ensuremath{{\rm \,cm}^{-3}}}
\newcommand{\pyr}{\ensuremath{{\rm yr}^{-1}}}
\newcommand{\be}{\begin{equation}}
\newcommand{\ee}{\end{equation}}
\newcommand{\ra}{{\rightarrow}}

\shorttitle{HOT H$_2$ EMISSION IN DISKS}
\shortauthors{\'AD\'AMKOVICS, NAJITA \& GLASSGOLD}

\begin{document}

\title{FUV Irradiated Disk Atmospheres: Ly$\alpha$ and the Origin of Hot H$_2$ Emission}

\author{M\'at\'e \'Ad\'amkovics$^1$, 
		Joan R. Najita$^2$, and 
		Alfred E. Glassgold$^1$}
\affil{
$^1$Astronomy Department, 501 Campbell Hall, 
	University of California, Berkeley, CA 94720 \\
	mate@berkeley.edu and aglassgold@berkeley.edu \\
$^2$National Optical Astronomical Observatory, 
	950 North Cherry Avenue, Tucson, AZ 85719;
 	najita@noao.edu
}

\begin{abstract} 
Protoplanetary disks are strongly irradiated by a stellar FUV spectrum that is dominated by \Lya\
photons. We investigate the impact of stellar \Lya\ irradiation on the terrestrial planet region of
disks ($\lesssim 1$\,AU) using an updated thermal-chemical model of a disk atmosphere irradiated by
stellar FUV and X-rays. The radiative transfer of \Lya\ is implemented in a simple approach that
includes scattering by \hi\ and absorption by molecules and dust. Because of their non-radial
propagation path, scattered \Lya\ photons deposit their energy deeper in the disk atmosphere than
the radially propagating FUV continuum photons. We find that \Lya\ has a significant impact on the
thermal structure of the atmosphere. Photochemical heating produced by scattered \Lya\ photons
interacting with water vapor and OH leads to a layer of hot (1500 -- 2500\,K) molecular gas. The
temperature in the layer is high enough to thermally excite the \hm\ to vibrational levels from
which they can be fluoresced by \Lya\ to produce UV fluorescent \hm\ emission. The resulting
atmospheric structure may help explain the origin of UV fluorescent \hm\ that is commonly observed
from classical T~Tauri stars.
\end{abstract}


\slugcomment{Accepted to ApJ December 14, 2015}
			 
\section{INTRODUCTION}

Stellar FUV irradiation is expected to strongly affect the properties of protoplanetary disk
atmospheres  \citep[e.g.,][]{Aikawa1999, Willacy2000, Markwick2002, vanZadelhoff2003, Gorti2004,
Gorti2008, Kamp2004, Jonkheid2004, Nomura2007, Agundez2008, BB09, Ercolano2009, Woitke2009a,
Woods2009, Heinzeller2011, BB11, Fogel2011, Cleeves2011, Walsh2012, Akimkin2013, Du2014}. FUV
photons (912 -- 2000\AA) not only ionize atoms, dissociate molecules, and contribute to
photoelectric heating, but they also induce photochemical heating
\citep{BB09,GMN09,Woitke2009a,Gorti2011,AGN14,GN15}. The absorption of FUV photons in disk
atmospheres can shield the disk midplane from irradiation \citep{BB09, GMN09, AGN14}.

The stellar FUV radiation field has contributions from both the continuum and \Lya.  Valuable clues
to the strength of the \Lya\ component are provided by observations of UV fluorescent \hm\ emission
from T Tauri stars (TTS). The emission, which is well explained by strong dipole electronic
transitions pumped for the most part by \Lya\ photons from excited vibrational levels of \hm\
\citep{Ardila2002, Herczeg2002}, is detected commonly from the inner regions of protoplanetary
disks. In the {\it Hubble Space Telescope (HST)} Cosmic Origins  Spectrograph (COS) survey reported
by \citet{France2012}, fluorescent \hm\ emission is detected from all accreting young stars
(classical T Tauri stars; CTTS). Spectrally resolved line profiles of the \hm\ emission are
consistent with the bulk of the emission arising from a circumstellar disk at radii within an AU
\citep{France2012}. If the \hm\ absorption lines are optically thin, the observed
\hm\ fluorescent emission can be used to reconstruct the \Lya\ luminosity incident on the disk.
Studies that have used this approach find that \Lya\ is the dominant component ($\sim$80\%) of the
stellar FUV field \citep[e.g.,][]{Herczeg2004, Schindhelm2012}.

A TTS FUV field that is dominated by \Lya\ may impact the disk chemistry significantly
\citep{Bergin2003}. Several recent studies have included stellar \Lya\ as a distinct component of
the stellar FUV field. \citet{BB11} have explored in detail the effect of \hi\ resonant scattering
on \Lya\ transport in TTS disks that have experienced significant dust settling. Models that include
photodissociation by \Lya\ separately from the FUV continuum can bring model abundance ratios into
better agreement with observations of outer disks \citep{Fogel2011,Walsh2012}. Detailed and
sophisticated models have also explored the effect of \Lya\ on the water vapor distribution in disks
\citep{Du2014}.

\Lya\ may affect disk properties significantly not only because it is more luminous, but also
because of the different path that \Lya\ photons take in reaching the disk. As described by
\citet[][hereafter BB11]{BB11}, stellar \Lya\ photons are strongly scattered by \hi\ in the upper
disk atmosphere. Photons scattered downward take a much more direct path into the disk than stellar
continuum photons, which propagate through the disk at an oblique angle. The latter traverse a
column of material along the line of sight from the star to a given depth into the disk that is much
larger than the vertical column, and are likely to be strongly absorbed higher in the atmosphere. As
a result, \Lya\ is expected to dominate the UV photon field at large vertical column densities.

Despite the significance of these inferences from UV fluorescent \hm\ emission, the origin of the
disk conditions that produce the UV \hm\ emission have been difficult to explain. Populating the
excited vibrational levels of \hm\ that can couple to the \Lya\ field requires a gas temperature
that is unusually high for molecular gas ($\sim$2500\,K) if the excitation is thermal
\citep{Herczeg2004, Schindhelm2012}. Since these temperatures are larger than have been expected for
\hm\ in protoplanetary disk atmospheres \citep[e.g.,][]{Woitke2009a,Walsh2012,GNI04}, it is
often assumed that the emitting \hm\ is populated nonthermally, by UV and X-ray emission, as in the
model of \citet{Nomura2007}. However, if there is a mechanism for heating the molecular gas 
to $\sim$2500\,K, then the excited vibrational levels of \hm\ can be populated thermally.

Here we revisit the problem of the origin of hot \hm\ emission from disks using an expanded
thermal-chemical model of protoplanetary disk atmospheres that explores the role of \Lya\ in heating
the atmosphere. The earlier version of our model implemented irradiation by stellar FUV continuum
photons and a preliminary treatment of photochemical heating by \water\ and OH \citep[][hereafter
AGN14]{AGN14}. In the present study, we include irradiation by stellar \Lya, self-shielding in the
900--100\AA\ band, and an improved implementation of photochemical heating. The \Lya\ radiative
transfer is treated in a schematic way.

We find that \Lya\ irradiation is an important heat source for disk atmospheres. Our model can
account for the characteristic properties of the hot \hm: its characteristic temperature, column
density, and emitting radii. Our model and the updates to it are described in Section~2. In
Section~3 we demonstrate the role of \Lya\ heating in determining the thermal and chemical structure
of the atmosphere. These results are discussed in Section~4 and we summarize our findings in
Section~5.

\section{\label{s:model}THERMAL-CHEMICAL MODEL}

We use a thermal-chemical model of an X-ray and FUV irradiated disk that was most recently described
in AGN14. The model builds on the work presented in \citet{GNI04},
\citet{GMN09} and \citet{AGM11}. As described in our previous work, the disk model atmosphere has a
layered structure, with a hot ($\sim$5000\,K) atomic layer overlying warm ($\sim$800\,K) molecular
and cool ($\sim 500$\,K) molecular layers. The resulting properties of the model atmosphere, such as
warm columns of species and their radial extent, are in good agreement with the general molecular
emission properties of {\it Spitzer} spectra of protoplanetary disks \citep[][AGN14]{NAG11}.

The model adopts a static disk density and dust temperature structure from  \citet{DAlessio1999},
with stellar and disk parameters listed in Table~\ref{t:std}. As in AGN14, we assume dust properties
(Table~\ref{t:std}) that take into account that large grains settle to the midplane and leave a
reduced population of small grains in the atmosphere. The grain size parameter \ag\ corresponds to
a decrease in grain surface area by a factor of 20 compared to interstellar conditions, and a dust
surface area per hydrogen nucleus that is $\Sd\approx 8\times10^{-23}\sqcm$. A set of thermal and
chemical rate equations is used to determine the gas temperature and species abundances. The primary
improvements to the model are in the treatment of FUV radiation and a more complete set of
photodissocation pathways that use the stellar FUV field to heat the gas. Whereas we had previously
included the photodissociation of water and OH (AGN14), here we also include the FUV photochemistry
of additional abundant molecules and atoms and adopt the photochemical heating rates detailed in
\citet[][hereafter GN15]{GN15}. Most notably, we  now consider the radiative transfer and
photochemistry of \Lya\ as separate from the FUV continuum.

\begin{deluxetable}{lcl}
\tablecaption{\label{t:std}Reference Model Parameters}
\tablehead{
Parameter & Symbol & Value}  
\startdata 
Stellar mass           & $M_*$    &  0.5 $\MSun$   \\
Stellar radius         & $R_*$    &  2 $\RSun$     \\
Stellar temperature    & $T_*$    &  4000\,K       \\
Disk mass              & $M_D$    &  0.005 $\MSun$ \\
Disk accretion rate    & $\dot M$ &  10$^{-8}\Msun\,\pyr$\\
Dust to gas ratio      & $\rhod / \rhog$  & 0.01 \\
Dust grain size        & $\ag$    &  0.7 \micron   \\
Dust extinction        & $Q_{\rm ext}$ & 1.0 \\
X-ray temperature      & $\Tx$    &  1 keV         \\
X-ray luminosity       & $\Lx$    &  2 $\times 10^{30}\, \ergps$ \\
FUV continuum luminosity\tablenotemark{a} & \LFUV  &  1 $\times 10^{31}\, \ergps$ \\
\Lya/FUV continuum\tablenotemark{b}& $ \eta $  &  3 \\
Accretion heating      & $\ah $   &  0.5
\enddata
\tablenotetext{a}{The FUV continuum luminosity is integrated from 1100--2000\,\AA\ and excludes 
                  \Lya, so that it is smaller than the value used in AGN14.}
\tablenotetext{b}{The ratio of the unattenuated downward \Lya\ photon number flux to 
                  the radially-propogating FUV continuum number flux in 1200-1700\AA\ band.}
\end{deluxetable}

In order to explore the effect of \Lya\ radiation on the disk atmosphere, we adopt a schematic
treatment of its radiative transfer (scattering and absorption) that is based on detailed work in
the literature, as described below. Scattering of FUV continuum photons (mostly forward) by grains
is not considered. While we find that the propagation path of \Lya\ has a significant impact on the
properties of the atmosphere (Section~\ref{s:Results}), our assumptions suggest that these results
should be considered illustrative rather than quantitative. An improved FUV radiative transfer that
includes scattering effects in a more realistic way is needed to understand the effect of \Lya\ on
the detailed properties of the atmosphere.

The model presented here suffers from some additional shortcomings. Gas pressure balance is not
enforced in that our calculation of the gas temperature does not alter the density structure of the
atmosphere, as in the \citet{DAlessio1999} model. We effectively assume that dust heating dominates
in determining the density structure of the atmosphere. Because the gas temperature exceeds that of
the dust in the atmosphere, we therefore tend to overestimate the density. We further assume that
the dust abundance and size distribution are the same at all heights and radii in the atmosphere.
Transport of material (radial or vertical) is also not considered, and we assume that the chemical
timescales are rapid enough that abundances and thermal rates are determined  by local conditions.

These effects are also likely to be more prominent at smaller disk radii. This is particularly
relevant as the dust is important for attenuating the FUV, for \hm\ formation, and for thermal
accommodation of the gas at high densities. Detailed consideration of changes in the dust profile
with height, radius, and time are all very interesting and will be the subject of future work. In
particular, at radii within $\sim$0.1\,AU the dust structure of the inner rim may be important for
determining the radiation environment in the terrestrial planet region. For now, however, we
describe a simplified treatment below.

\subsection{FUV Radiative Transfer and Photochemistry}

Previous work in the literature serves as a valuable guide to understanding the propagation of \Lya\
in disk atmospheres. Studies have found that \Lya\ emitted by the star can be scattered by
intervening \hi, e.g., in a wind, before it reaches the disk \citep{Herczeg2004,Schindhelm2012}.
\hi\ in the atomic layer at the disk surface will also scatter arriving \Lya\ photons so that they
emerge roughly perpendicular to the disk surface (BB11). While a large fraction of the stellar \Lya\
photons may, in this way, be scattered away (reflected) from the disk, the fraction that passes
through the atomic layer travels a path more directly downward into the disk, in contrast to the FUV
continuum photons, which propagate at an oblique angle into the disk.

Quantitative results obtained in the earlier studies suggest a simple way to implement \Lya\
irradiation in our model. In their detailed Monte Carlo calculation of how \Lya\ photons interact
with the gaseous disk, BB11 start out with a number flux of \Lya\ photons 6 times larger than the
FUV continuum flux at the star. Upon leaving the star, \Lya\ photons are scattered away by
intervening \hi\ so that at the transition from atomic to molecular conditions in the disk
atmosphere, the \Lya\ flux is roughly equal to flux of the (obliquely propagating) FUV continuum
photons. A roughly similar ratio of \Lya\ to continuum photons is inferred observationally for the
disk surface. In their analysis of the \hm\ fluorescence emission from CTTS, \citet{Schindhelm2012}
found that the flux of \Lya\ incident on the fluorescing \hm\ layer (their $F_{ab}$) is 1.5--5 times
larger than the continuum flux reaching the disk. Since some fraction of the stellar \Lya\ is
absorbed or scattered away before reaching the fluorescing \hm, the \Lya\ flux incident on the
\hm\ is less than the stellar \Lya\ flux.

We therefore assume for our reference model that at the top of the atmosphere the ratio of the
number flux of downward-propagating \Lya\ photons to the number flux of FUV continuum photons in the
1200--1700\AA\ band is $\eta=3$, a value in the middle of the range reported by
\citet{Schindhelm2012}. We apply the methods described in AGN14 and assume a mean FUV continuum
luminosity per 100\,\AA\ of $L_{\mathrm{band}} = 1.1\times 10^{30}\,$erg~s$^{-1}$. The FUV
luminosities of CTTS can be an order of magnitude larger or smaller \citep{Yang2012}. The FUV
continuum is absorbed by molecules and dust along the line of sight to the star. We improve on the
treatment of the FUV continuum photoabsorption presented in AGN14 by treating a larger number of
abundant molecules and atoms: \hm, CO, \nm, C, O$_2$, NH$_3$, HCN, CH$_4$,  C$_2$H$_2$, and SO$_2$
in addition to H$_2$O and OH.

\Lya\ photons are also attenuated by dust and molecules, and we incorporate, in an approximate way,
scattering of \Lya\ by \hi\ in the atmosphere. The scattering and absorption cross sections
illustrate the relative roles of these processes for \Lya. The UV dust absorption cross section is
$8\times 10^{-23} \sqcm$ per hydrogen, while the \water\ absorption cross section at \Lya\ is
$\sigma_{\rm a}(\water) \approx 10^{-17}\sqcm$ \citep{vanDishoeck2006}. For a typical water
abundance of $x(\water) \approx 10^{-4}$,
\be
\frac{\sigma_{\rm a}({\rm dust})}{x(\water) \sigma_{\rm a}(\water)} \approx 0.08.
\ee
Thus, water will typically dominate the UV opacity in the molecular layer. 

To compare the roles of \water\ absorption and \hi\ scattering for \Lya, we can consider a
representative \hi\ scattering cross-section.  At $\sim$80 Doppler widths ($\sim$300 km/s) from
\Lya\ line center, where the \Lya\ line profile peaks in the spectrum of the CTTS TW~Hya
\citep{Herczeg2004}, the \hi\ scattering cross section at that velocity is $\sigma_{\rm
s}(\h)\approx10^{-20}\sqcm$ (BB11). At the top of the molecular layer 
$\xh/x(\water)$ ranges from $3\times 10^5$ to 10, 
so that the ratio
of \hi\ scattering to \water\ absorption optical depths 
\be \label{e:HI_scat_approx}
\frac{\xh \sigma_{\rm s}(\h)}{x(\water) \sigma_{\rm a}(\water)} \approx 300-0.01.
\ee
The ratio in Eq.~\ref{e:HI_scat_approx} is a sensitive function of depth into the 
atmosphere, because $x(\h)$ and $x(\water)$ both depend strongly on vertical column. This 
comparison suggests that there may be regions where  \hi\ scattering can increase the path of 
\Lya\ photons through the absorbing medium, and can therefore increase the probability of 
absorption by \water.

To include \Lya\ absorption and scattering in a simple way, we can consider the scattering optical
depth along a path length $\ell$ through a grid cell $ \tau_{\rm s} = n_{\rm s} \sigma_{\rm s}
\ell = \ell/\ell_{\rm s}$, where $n_{\rm s}$ and $\sigma_{\rm s}$ are the number density and cross
section of scatterers, and $\ell_{\rm s}$ is the scattering mean free path. To traverse a distance
$\ell$ via a random walk, the \Lya\ photons will take approximately $\tau_{\rm s}^2$ steps. That is,
scattered photons will travel a distance $\ell_{\rm eff} = \tau_{\rm s}^2 \ell_s = \tau_{\rm s}\ell
$ through the absorbing medium. Because of the longer path length, the effective optical depth for
absorption is
\be
\label{e:tau_eff}
	\tau_{\rm eff} = n_{\rm a} \sigma_{\rm a} \ell_{\rm eff}  
      	= (n_{\rm a} \sigma_{\rm a} \ell)\tau_{\rm s} 
	= \tau_{\rm a} \tau_{\rm s}
\ee
where $n_a$ and $\sigma_a$ are the number density and absorption cross section of absorbers (e.g.,
dust or molecules) and $\tau_{\rm a} = n_{\rm a} \sigma_{\rm a} \ell$ is the usual absorption
optical depth without scattering. As a result, for each grid step in the model, we calculate 
the scattering optical depth as
\be
\tau_{\rm s} = n(\h) \sigma_{\rm s}(\h)\ell_z,
\ee
where $\ell_z$ is the vertical size of the grid cell, and approximate the increased probability of
absorption in the presence of scattering by multiplying the normal absorption optical depth
$\tau_{\rm a}$ by a factor of $\tau_{\rm s}$ when the scattering is large (i.e., $\tau_{\rm s} >
1$). The increase in path length leads to increased \Lya\ absorption, which both attenuates the
\Lya\, and increases the photochemical and photoelectric heating by \Lya. We also assume that
fluorescent excitation of hot \hm\ does not attenuate the \Lya, consistent with detailed
modeling of UV fluorescent H2 emission, which finds that only a small fraction of the incident
\Lya\ is processed into \hm\ emission (2\% for TW~Hya; Herczeg et al.\ 2004). The photodissociation
rate for species X is given as the sum of the dissociation rates from FUV continuum and \Lya\
photons,
\be \label{e:Gx} 
G({\rm X}) = \Gcnt + \GLya.
\ee 
The continuum rate is
\be \label{e:GFUV} 
\Gcnt = \int_{\lambda_0}^{\lambda_f}  \Fcnt(\lambda) \sigma(\lambda ; {\rm X}) d\lambda,
\ee
where $\Fcnt(\lambda)$ is the local FUV continuum photon number flux spectrum over wavelengths
$\lambda$, which has been attenuated along the line of sight to the star and $\sigma(\lambda ; {\rm
X})$ is the photodissociation cross section spectrum for species X. Similarly, the photorate for
dissociation by \Lya\ is
\be
G_{\Lya} = F_{\Lya}\,\sigma(\Lya ; {\rm X}), 
\ee 
where $F_{\Lya}$ is the downward propagating \Lya\ number flux, and $\sigma(\Lya ; {\rm X})$ is the
photodissociation cross section for species X at \Lya. References for the cross sections used here
are given in Table~\ref{t:phchem}.

The opacity of the 911 -- 1108\,\AA\ band is dominated  by the abundant species \hm, CO, \nm, and C.
Most importantly, \hm, CO and \nm\ absorb via lines and thus self- and mutually shield one another,
as well as all other species. They therefore require detailed treatment of their opacity. Being the
most abundant, \hm \, is the most important absorber in the 911 -- 1108\,\AA\ band. We use the \hm\
shielding function given by \citet{DB96},
\be
\label{e:jH2}
J(\hm) = \frac{0.965}{(1+x/b_5)^{\alpha}}+\frac{0.035}{\sqrt{1+x}}
		  e^{-8.5\times10^{-4} \sqrt{1+x}},
\ee
where $x \equiv N(\hm)/5\times 10^{14}\,\psqcm$, $b_5 \equiv b/10^5\cm\ps$, and the Doppler
broadening parameter is $b=3$\,km\,\ps. The original expression used $\alpha$=2 and applied to
temperatures up to 300\,K. \citet{Wolcott-Green2011} determined the shielding function in a 3D model
and considered temperatures up to 10$^4$\,K. They found that using $\alpha=1.1$ gives a better
parameterization of the shielding at high temperatures, which we use here in calculating $J(\hm)$.

Tabulations of the shielding by CO as a function of $N(\hm)$ and $N({\rm CO})$ are provided by
\citet{Visser2009} for a set of Doppler widths, excitations temperatures, and isotopologue
ratios\footnote{\href{http://home.strw.leidenuniv.nl/~ewine/photo/}
{http://home.strw.leidenuniv.nl/$\sim$ewine/photo/} } . We use the table for the conditions that
most closely approximate the conditions in the molecular region of the disk, $b({\rm
CO})=0.3$\,km\ps, $T_{\rm ex}({\rm CO})=100$\,K, and $^{12}$CO/$^{13}$CO=69, to lookup the CO
shielding function $J({\rm CO})$. Similarly, \citet{Li2013} provide the shielding functions for
\nm, and we use the shielding functions $J({\rm N_2})$ tabulated for 
$b({\hm})=\,3.0$\,km\ps, $b({\nm})=\,0.77$\,km\ps, $T_{\rm ex}({\rm N_2})=T_{\rm ex}({\rm
H_2})=1000$\,K, and $N(\h)=10^{22}$\psqcm. 

For \hm, CO, and N$_2$, the first (continuum) term in the expression for $G({\rm X})$ is decreased
according to the shielding factors described above, as well as by all other continuum absorbers
including dust grains. The absorption spectrum of C in the 911 -- 1108\,\AA\ band is essentially
continuum absorption, and for large \Nh, the molecular hydrogen shielding occurs in the far line
wing, which covers a significant fraction of continuum. Therefore, \hm\ may shield the ionization of
atomic carbon, and so we estimate the attenuation of $G({\rm C})$ by a factor of $J(\hm)$, defined
in Eq.~\ref{e:jH2}.

\subsection{\label{s:Compton}X-rays with Compton Scattering}

The theory for X-ray ionization presented in \citet{AGM11} and implemented in AGN14, uses a single
temperature X-ray spectrum, and considers only the absorption of X-rays. Studies by \citet{EG13},
using a 3D radiative transfer and photoionization code with Compton scattering together with a more
realistic two-temperature X-ray spectrum, show that ionization rates can be factors of several
larger  at low densities (and smaller at high densities) in the disk atmosphere than in our earlier
calculations. \citet{EG13} tabulate ionization rates calculated with depleted ISM elemental
abundances for X-ray spectra that match the observations of the {\it Chandra} Orion Ultradeep
Project (COUP). Since we consider the same disk density structure, we scale total ionization rates
in \citet{AGM11} to match \citet{EG13} at each altitude in the model, and we use this scaling to
calculate the shell-specific ionization rates.

\subsection{Thermal Processes: Photochemical heating by \Lya}

Our thermal model includes the heating processes described in \citet{GNI04} and AGN14: X-ray
heating, accretion-related mechanical heating,  thermal accommodation between gas and dust,  grain
photoelectric heating,  photochemical heating by \water\ and OH,  and \hm\ formation heating. In
addition we include photochemical heating  for C, \hm, CO, \water, and OH following GN15, as well as
O$_2$, which is described in the Appendix. We do not calculate the photochemical heating for NH$_3$,
HCN, CH$_4$,  C$_2$H$_2$, and SO$_2$ due to their small abundances. We include photochemical heating
for \water\ and OH by \Lya\ photons, because these two molecules dominate the molecular opacity at
\Lya. The photochemical heating from other species, which do not contribute significantly to the
\Lya\ opacity, are ignored. Line cooling is essentially the same as in our earlier work and includes
\hi\ recombination lines, Ly$\alpha$, \hm\ vibrational and rotational transitions,  CO rovibrational
and pure rotational transitions, \water\ vibrational and rotational transitions, and  \oi\ forbidden
and fine structure transitions.  Above the atomic to molecular transition,  \Lya\ cooling dominates
and is supplemented at the transition by CO rovibrational, dust-gas and  \oi\ forbidden-line
cooling.

In the prescription given in AGN14 for the chemical heating associated with the photodissociation of
water and OH, roughly half of the photon energy in excess of the dissociation energy was converted
to heating, primarily through collisions of the dissociation products. GN15 provide a more detailed
description of the energetics of the dissociation products  of these and other molecules and the
thermal energy that could potentially be produced in their subsequent chemical reactions. The total
heating due to the photodissociation of species X, $Q({\rm X})$, is the sum of the  direct heating
component, $Q_{\rm dir}({\rm X})$, which arises from the translational energy of the dissociation
products, and the chemical heating component, $Q_{\rm  chem}({\rm X})$, which arises from the
excitation of the products and their subsequent chemical reactions. The heating rate per
photodissociation of each species is the sum of the heating by FUV continuum photons, $Q_{\rm
cont}({\rm X})$, and by \Lya\ photons, $Q_{\Lya}({\rm X})$,
\be 
\Gamma_{\rm phchem} =  \Gcnt \, n \, Q_{\rm cont} +  G_{\Lya} \, n \, Q_{\Lya}, 
\ee 
where $n$ is the volumetric number density of a particular species, having dropped the X from the 
notation. 

Water and OH are the dominant sources of opacity for \Lya\ in the molecular layer as well as the
dominant sources of photochemical heating. The photodissociation of \water\ by \Lya\ has three
possible product channels \citep{Harich2000}, which were used in GN15 to calculate FUV continuum
dissociation at wavelengths below 1450\AA\ (their Band 2). In this band, GN15 obtained direct and
chemical heating energies of $Q_{\rm dir}$\,=\,0.24 eV and $Q_{\rm chem}$\,=\,0.86 eV, respectively,
for very high densities. We can apply the continuum results of GN15 that were based on experiments
for \Lya\ photodissociation by simply raising the mean energy used in the GN15 treatment (9.67\,eV),
to the \Lya\ photon energy of 10.2\,eV. For \Lya\ dissociation of \water, $Q_{\rm dir}$ becomes
0.8\,eV and the total photochemical heating is $Q_{\Lya}(\water)=Q_{\rm dir}+Q_{\rm chem}$=1.6\,eV.
Similarly, we can recalculate the heating per OH dissociation in GN15 using the 10.2\,eV \Lya\
photon instead of the mean FUV continuum photon energy. Taking into account the 4.4\,eV dissociation
energy of OH and the 2.0\,eV excitation energy of the excited O($^1$D$_2$) photoproduct, we have
3.8\,eV of direct heating energy. The GN15 chemical heating energy is 2.6\,eV, so that
$Q_{\Lya}(\oh)$=6.4\,eV. The photochemical heating energies for FUV continuum and \Lya\ are listed
in Table~\ref{t:phchem}.

\section{RESULTS}
\label{s:Results}

\begin{figure}[]\begin{center}
\includegraphics[width=3.25in]{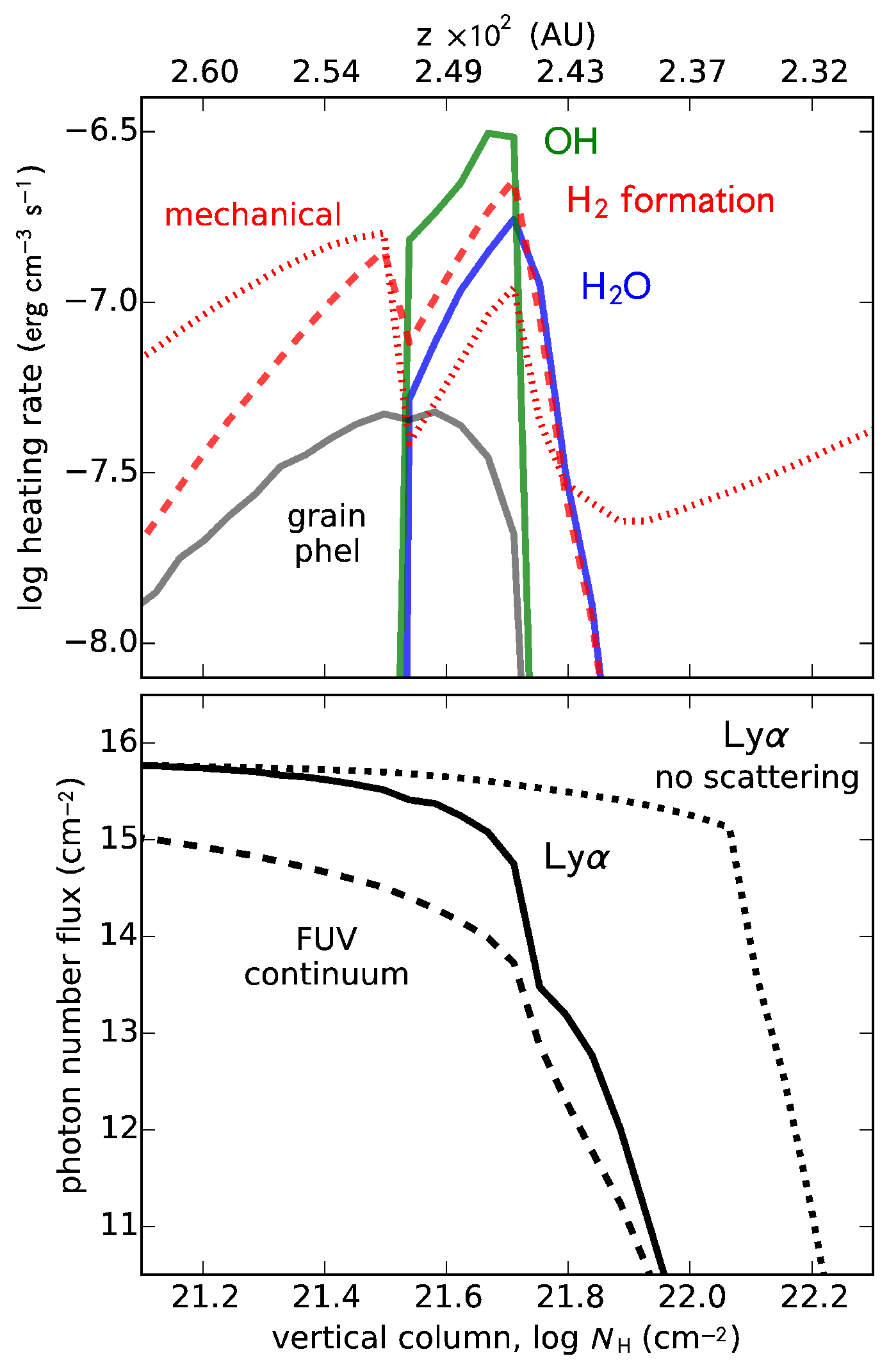}
\caption{\label{f:FUV_heat} Vertical profiles of the thermal rates at $r=$0.24\,AU for the dominant
heating mechanisms are shown in color in the top panel. The FUV heating is dominated by the
photodissociation of molecules, primarily \water\ and OH by \Lya\ (blue and green curves,
respectively), along with the closely related heating due to \hm\ formation on grains (red dashed
curve). Photoelectric heating on dust (grey curve) and mechanical heating (red dotted curve) are
important in the atomic region at vertical columns less than $\log\Nh\approx 21$. Vertical profiles
of the photon number fluxes (bottom panel) for the FUV continuum  (black dashed curve), \Lya\ in the
absence of scattering (black dotted curve), and \Lya\ including the additional opacity due to
scattering described in Equation~\ref{e:tau_eff} (black solid curve). A physical distance scale for
the altitude above the midplane is given on the top axis.}
\end{center} \end{figure}

We calculate the vertical (altitude) structure of abundances $x$, and gas temperature \Tg, at radial
distances of 0.24, 0.48, and 0.95\,AU, which are characteristic of the emitting regions inferred
from \hm\ line widths in CTTS \citep[][Table~4]{France2012}. Although many thermal processes are
considered here, in general only a few play major roles in the disk atmosphere. For example, in the
low density layer just above the atomic to molecular transition, accretion heating is balanced by
\Lya\ cooling, but cooling by CO rovibrational emission and \hm\ formation heating are also
important. The dominant heating mechanisms are plotted as a function of the vertical column density
of hydrogen from the top of the disk atmosphere  \Nh\ in Figure~\ref{f:FUV_heat}, along with the
\Lya\ and FUV continuum fluxes, for a radial distance of $r=0.24$\,AU.

As shown in Figure~\ref{f:FUV_heat}, heating by \water, OH, and \hm\ formation dominates other
heating mechanisms at the top of molecular layer ($\log N_H = 21.5-21.7 \psqcm$) and raises the
temperature of the molecular gas above 1500\,K. These heating processes can each exceed mechanical
and photoelectric heating. The \hm\ formation heating in the warm molecular layer is actually
photochemical in origin, as discussed in GN15. Deeper into the molecular layer the FUV radiation
(both continuum and \Lya) is shielded by both dust and molecules and neither mechanism has an
important role in heating the atmosphere.

\begin{figure}[t]\begin{center}
\includegraphics[width=3.25in]{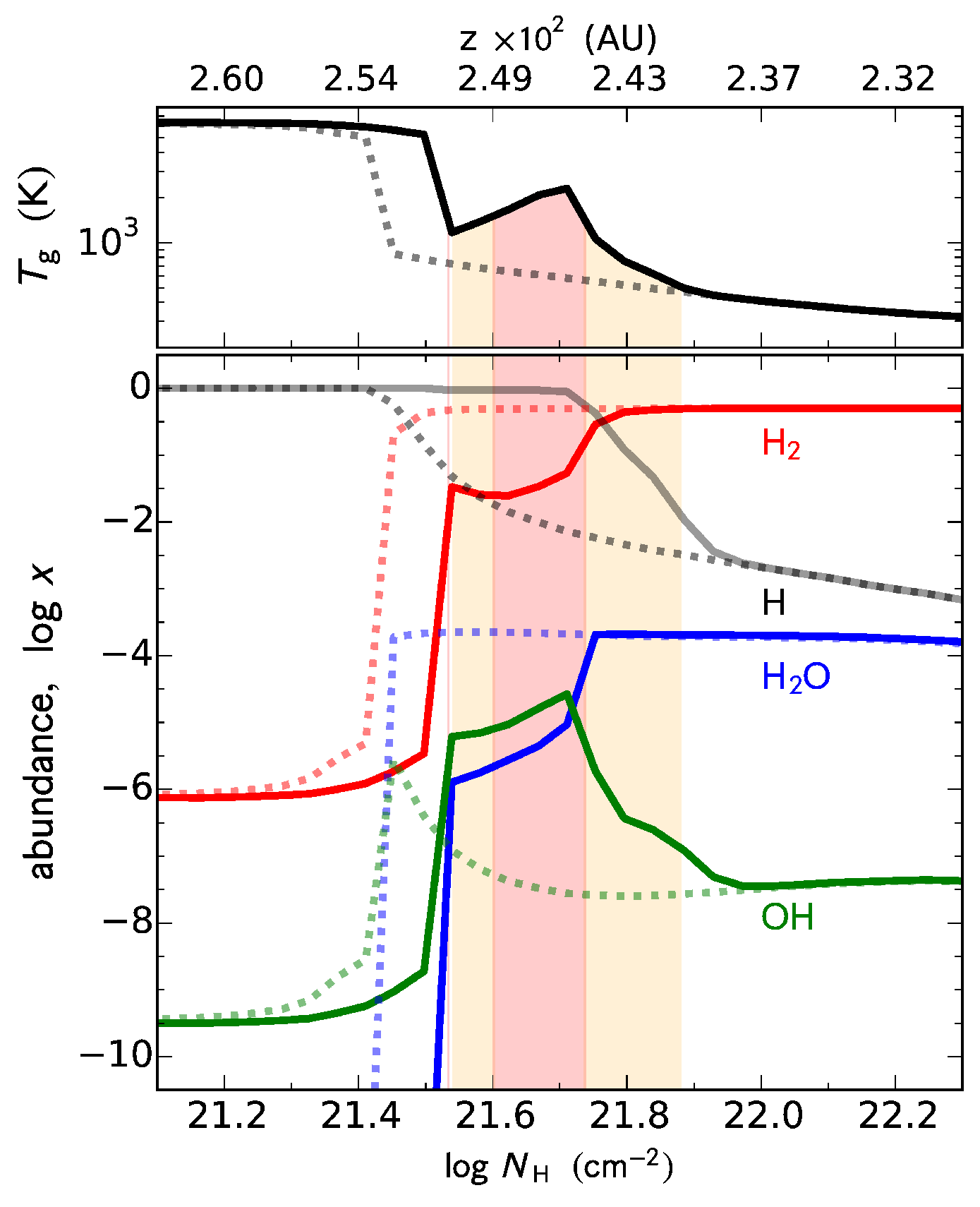}
\caption{\label{f:molecules}Vertical profiles of molecular abundances and gas temperature in the
reference model of the disk atmosphere at $r=0.24$\,AU  (solid curves). The red shaded region
highlights the location of gas that is sufficiently hot for UV fluorescence from \hm, where \Tg\ is
in the range 1500 -- 2500\,K. The orange shaded region is where \Tg\ is in the range 500 -- 1500\,K.
Dotted curves are from a model that has no \Lya. A physical distance scale for the altitude above
the midplane is given on the top axis.}
\end{center} \end{figure}

Figure~\ref{f:molecules} shows the vertical profiles of key molecular abundances and temperatures
plotted for the reference model (solid curves), and for comparison we also show the results for a
model without \Lya\ (dotted curves), which are essentially the same as our earlier reference model
(i.e., the top left panel of Figure~5 in AGN14). In the absence of \Lya, there is a steep transition
to peak \xhm\ and peak $x(\water)$ near $\log\Nh=21.4\psqcm$ and after the transition there is a
steady decline in gas temperature and \xoh\ with increasing \Nh. Thus as in earlier models, the disk
atmosphere is characterized by a hot ($\sim$5000\,K) atomic layer that overlies a cooler
($\sim$1000\,K) molecular layer. The region between these two layers is where the role of FUV
radiation and photochemistry is most important, with the strength of the radiation field determining
whether there is a sharp or a gradual transition from atomic to molecular conditions (AGN14).
 
As illustrated by our reference model with \Lya\ (solid curves in Figure~\ref{f:molecules}), 
\Lya\ radiation is important for both heating the gas and photodissociating molecules near the
transition. With \Lya\ included, the transition occurs deeper than with FUV continuum radiation
alone, and e.g., the abundance of molecules such as \hm\ and \water\ are reduced near
$\log\Nh\approx21.6\psqcm$. This is similar to the effect of increasing the FUV continuum
luminosity when \Lya\ is absent (Figure~4 in AGN14). Since the vertical optical depth of the
atmosphere is roughly an order of magnitude less than the line of sight optical depth, the downward
propagating \Lya\ penetrates deeper into regions of higher density than the FUV continuum, which
travels along an oblique line-of-sight from the star. In contrast to our models without \Lya, the
gas temperature \Tg\ in our reference model increases with vertical column \Nh\ after the
transition, from $\log\Nh$ = 21.5 through 21.7 \psqcm\ as a result of photochemical heating by \Lya.
The red shaded region in Figure~\ref{f:molecules} highlights the layer of hot molecular gas, where
\Tg\ is 1500 -- 2500\,K. The total column of \hm\ in this region is 4.9$\times 10^{19}\psqcm$. In
the absence of \Lya\ the gas temperature is much lower, and there is no region of the atmosphere
where \hm\ is both abundant and hot.

The depth to which the FUV penetrates determines where hot molecular gas is present. As a result,
the total column of hot molecular gas is sensitive to the magnitude of the FUV. Since we assume that
the downward \Lya\ number flux is a factor of $\eta=3$ larger than the FUV continuum number flux at
the top of the atmosphere (see Section~2.1), an order of magnitude increase in the FUV continuum
leads to an order of magnitude more \Lya\ and produces a factor of $\sim$4 times more hot \hm\
(Table~\ref{t:H2}) at 0.24\,AU. Increasing \Lya\ also leads to higher temperatures at larger radii.
While there is no significant column of hot \hm\ in the reference model at $r=$0.48\,AU (with
$\LFUV=10^{31}\erg\ps$), and order of magnitude increase in $F_{\Lya}$ leads to significant columns
($>$10$^{19}$\psqcm) of hot gas out to 0.48\,AU in the disk.

\begin{deluxetable*}{l*{6}c}
\tablecaption{\label{t:H2}Hot H$_2$ Column Densities\tablenotemark{a} and Model Parameters}
\tablehead{
Calculation & \multicolumn{3}{c}{Radius (AU)} & 
              \multicolumn{3}{c}{Parameters\tablenotemark{b}} \\
            & 0.24 & 0.48  & 0.95 & \ah & \ag & \LFUV }  
\startdata 
                        & 3.6(15)      & 3.1(15)      & 3.8(15) &  0.50 &  0.71      & 1(30) \\
Reference\tablenotemark{c} \ah, \ag & {\bf4.9(19)} & 3.2(15) & 3.6(15) & 0.50 & 0.71 & 1(31) \\
                        & {\bf2.0(20)} & {\bf3.6(19)} & 4.5(15) &  0.50 &  0.71    & 1(32) \\ \hline
                        & 6.4(15) & 1.1(17) & 1.2(17) &  0.01 &  0.71                & 1(30) \\
Reduced \ah             & {\bf7.3(19)} & 6.0(15) & 8.0(15) &  0.01 &  0.71           & 1(31) \\
                        & {\bf1.9(20)} & {\bf5.0(19)} & 3.5(15) &  0.01 &  0.71    & 1(32) \\ \hline
                        & 1.2(15) & 2.2(15) & 1.6(15) &  0.50 &  7.07                & 1(30) \\
Larger \ag              & 1.4(15) & 1.6(15) & 2.0(15) &  0.50 &  7.07                & 1(31) \\
                        & {\bf3.4(20)} & {\bf9.9(18)} & 2.4(15) &  0.50 &  7.07    & 1(32) \\ \hline
                        & 1.1(15) & 2.5(15) & 1.9(16) &  0.01 &  7.07 & 1(30)                \\
Reduced \ah, larger \ag & {\bf3.9(19)} & {\bf5.8(19)} & 2.5(16) &  0.01 &  7.07 & 1(31)      \\
               & {\bf9.5(19)} & {\bf3.8(19)} & {\bf3.3(19)} &  0.01 &  7.07 & 1(32) \\ \hline \hline
No \Lya\                & 3.7(15) & 2.9(15) & 3.6(15) &  0.50 &  0.71 & 1(31)                \\ 
$\eta=1$                & {\bf2.8(19)} & 3.1(15) & 3.4(15) &  0.50 &  0.71 & 1(31)           \\
$\eta=3$\tablenotemark{c} & {\bf4.9(19)} & 3.2(15) & 3.6(15) &  0.50 &  0.71 & 1(31)         \\
$\eta=5$                & {\bf5.7(19)} & {\bf2.4(19)} & 3.7(15) &  0.50 &  0.71 & 1(31)      \\                        
No \Lya\ scattering     & {\bf2.9(20)} & {\bf1.7(20)} & 3.6(15) &  0.50 &  0.71 & 1(31)      \\
\enddata
\tablenotetext{a}{Column densities for gas temperatures $T>$1500\,K in units of \psqcm, with values 
                  above $10^{18}\psqcm$, which indicate a hot \hm\ region, highlighted in bold.}
\tablenotetext{b}{Parameter units: \ah\ is the dimensionless accretion-related mechanical heating 
                  parameter; \ag\ is the grain size parameter in \um; and \LFUV\ is the FUV 
                  continuum lumimosity in \erg\ps.}
\tablenotetext{c}{The reference model that is plotted for $r=0.24$\,AU in the Figures.}
\end{deluxetable*}

On the other hand, a reduction in the FUV radiation has an even more dramatic effect on the column
of hot molecular gas. We varied the radiation by decreasing both the FUV continuum and \Lya; we also
removed \Lya\ entirely by setting $F_{\Lya}=0$. In either case, the column of hot \hm\ drops
significantly. In models with either \LFUV=10$^{30}$\,\erg\ps or no \Lya, the maximum
temperature in the irradiated molecular layer is below 1500\,K, and the column of hot
\hm\ is four decades smaller (Table~\ref{t:H2}). In these cases, the only region of the atmosphere
where $\Tg > 1500$\,K is in the hot atomic layer, where $\xhm \lesssim 10^{-5}$.  As a result of the
small abundance of \hm, the total column of hot \hm\ in the atomic layer is only $\sim
10^{15}$\psqcm. This is the characteristic minimum value in Table~\ref{t:H2} that indicates the lack
of a hot molecular region. While both FUV continuum and \Lya\ are important for heating, the
absence of a hot molecular region when \Lya\ is removed -- while the FUV continuum is maintained ---
suggests that \Lya\ plays a more important role than the continuum in producing hot molecular gas.
Again, this is because the downward propagating \Lya\ photons penetrate deeper than the continuum.

\begin{figure}[t]\begin{center}
\includegraphics[width=3.25in]{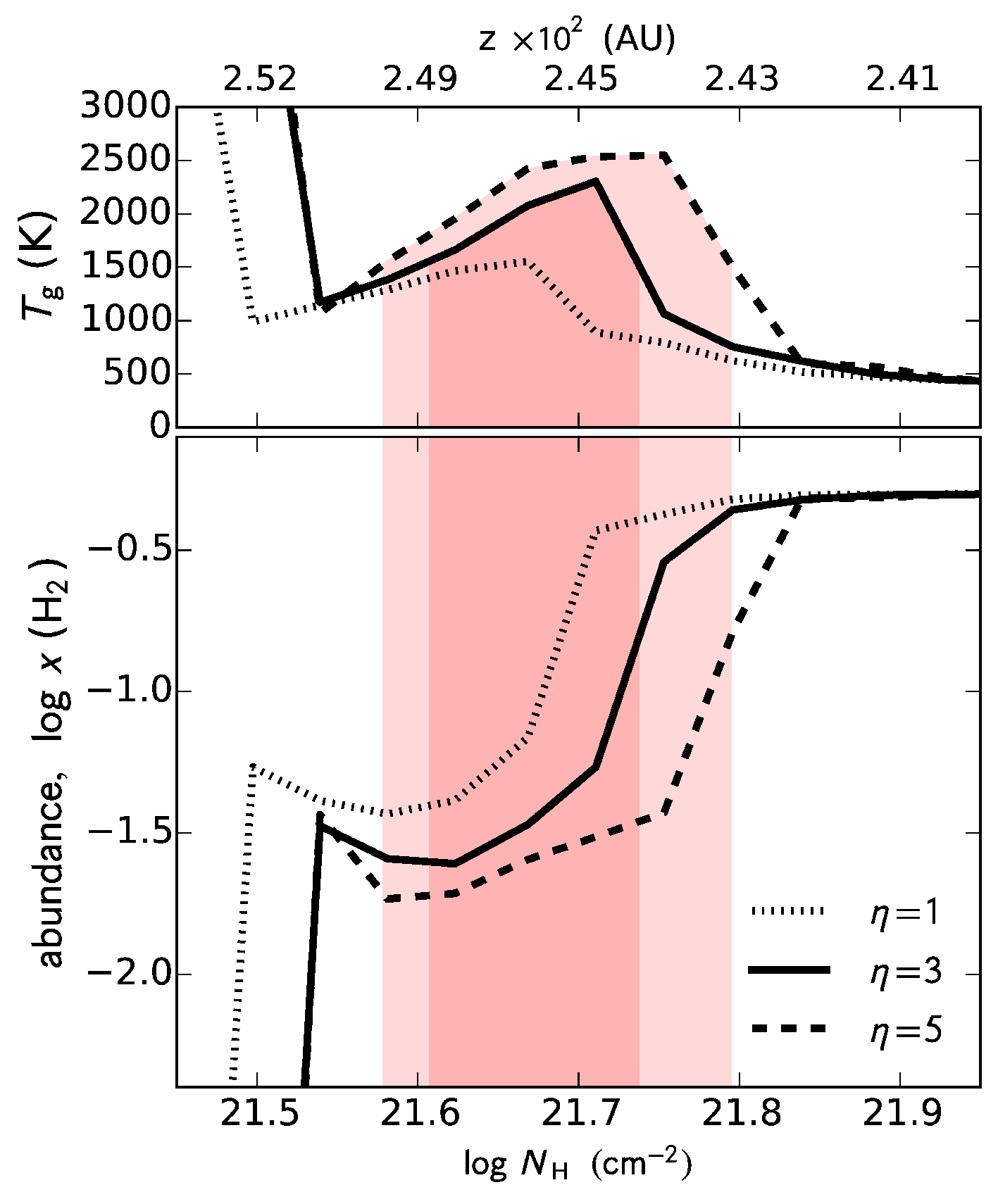}
\caption{\label{f:Lya_var}Vertical profiles of gas temperatures and \hm\ abundances for different
\Lya\ radiation fluxes at $r=0.24$\,AU, highlighting the role of \Lya\ in determining the extent of
the region of hot \hm. The reference model has $\eta=3$ (solid curves), and the region of hot \hm\
is highlighted in the darker shaded region. Also shown are models with $\eta=1$ (dotted), which
barely exceeds 1500\,K in the molecular layer, as well as models with $\eta=5$ (dashed), where both
the maximum \Tg\ and the depth of the hot \hm\ region are increased relative to the reference model
(lighter shaded region). A physical distance scale for the altitude above the midplane is given on
the top axis. }
\end{center} \end{figure}

The depth to which \Lya\ photons penetrate depends on scattering in the hot molecular region. To
illustrate the role of scattering we considered a case of pure absorption, i.e.,  only dust and
molecular absorption and no \hi\ scattering, which leads to the deepest penetration of radiation
(black dotted curve in the bottom panel of Figure~\ref{f:FUV_heat}) and nearly and order of
magnitude increase in the columns of hot \hm\ than when scattering is included (Table~\ref{t:H2}).
The column of hot \hm\ is also sensitive (obviously) to $\eta=F_{\Lya}/F_{\rm cont}$ at the top of
the atmosphere (Figure~\ref{f:Lya_var}). When \Lya\ is reduced by a factor of 3 relative to our
reference model (setting $\eta = 1$) the gas is heated to barely above 1500\,K. An increased value
of $\eta=5$ results in a higher peak temperature for the hot \hm\ and deeper penetration of the
radiation into the disk.

Two other parameters in the model that are generally important for heating are the accretion-related
mechanical heating parameter \ah, defined in \citet{GNI04}, and the dust grain size parameter, \ag.
Mechanical heating is the dominant heating term deep in the disk and in the hot atomic region, while
dust-gas thermalization is the dominant cooling mechanism at large column densities.
Table~\ref{t:H2}, which shows the hot \hm\ columns for cases with reduced \ah, larger \ag, or a
combination of the two, illustrates the sensitivity of our results to these parameters. Effectively
eliminating mechanical heating by reducing $\ah=0.01$ does not reduce the column of hot \hm\
(Table~\ref{t:H2}). This is because mechanical heating is subdominant to photochemical heating and
\hm\ formation heating in the region of hot \hm\ (Figure~\ref{f:FUV_heat}). Indeed, reducing \ah\
decreases the heating of the atomic layer, so that the transition from atomic to molecular
conditions occurs at a higher altitude in the disk atmosphere (Figure~\ref{f:par_var}). Models with
smaller \ah\ therefore have columns of hot \hm\ that are somewhat larger than the reference model,
by a factor of $\sim$50\% (Table~\ref{t:H2}).

The dust in the atmosphere plays competing roles as an important opacity source for the FUV, as the
catalyst for \hm\ formation and therefore \hm\ formation heating, and as an important coolant via
thermal accommodation in regions of high density. Increasing the grain size parameter, \ag, reduces
the surface area and opacity of the dust, reduces \hm\ formation heating, and allows the FUV to
penetrate deeper into the disk. With larger dust grains, the transition from atomic to molecular
conditions occurs at a lower altitude in the disk atmosphere, where the densities are higher
(AGN14). For the reference FUV continuum luminosity ($\LFUV=10^{30}\erg\ps$) and large grains
($\ag=7.07\um$), the transition is from hot atomic conditions to cool, fully molecular conditions,
and there is no region of hot \hm. However, models with increased radiation as well as with large
grains produce the largest columns of hot \hm\ (Table~\ref{t:H2}), as the irradiated molecular layer
occurs deeper into the disk (Figure~\ref{f:par_var}). The details of the dominant thermal and
chemical changes due to variations in \ah, \ag, and \LFUV\ are complex. However, as these parameter
variations illustrate, hot \hm\ columns $\sim 10^{19}\psqcm$ are a common outcome for a range
of \ag\ and \ah\ values.

\section{DISCUSSION}

We find that including \Lya\ in our earlier model of a disk atmosphere irradiated by FUV continuum
and X-rays (AGN14) produces a new component of the inner disk atmosphere: a region of hot molecular
gas (1500 -- 2500\,K; Figure~\ref{f:molecules}). The reason for the difference is related to the
distinctive radiative transfer and propagation path of the \Lya\ photons more than to their
luminosity. Although the \Lya\ component is much more luminous than the FUV continuum in the present
case, the total FUV luminosity (continuum + \Lya) is similar to that assumed in AGN14. More
significantly, the \Lya\ fraction of the FUV luminosity propagates more directly downward through
the disk compared to the FUV the continuum, which propagates along an oblique line-of-sight into the
disk. As shown in Table~\ref{t:H2}, a hot molecular component occurs under a wide range of
conditions. The column density of hot \hm\ is insensitive to the value of the mechanical heating
parameter \ah. Column densities of hot \hm\ $>10^{19}\psqcm$ are obtained for a range of
FUV luminosities, (i.e., FUV continuum luminosities $0.0025-0.025\,\Lsun$). Large column densities
are also found when both the grain size parameter and mechanical heating are reduced.

The FUV continuum luminosities we considered span the middle to upper range among TTS 
\citep{Yang2012} and are consistent with the properties of sources with well-studied hot \hm\
\citep{France2012, Schindhelm2012}. The properties of the hot \hm\ found in the models (temperature,
column density, radial distance) are similar to those inferred for the UV fluorescent molecular
hydrogen emission from T Tauri stars. For example, in their study of \hm\ emission from TW~Hya,
\citet{Herczeg2004} inferred an excitation temperature of $T_{\rm ex}$\,=\,2500 (+700/-500)\,K, an
\hm\ column density of $\log N(\hm) = 18.5 (+1.2/-0.8) \psqcm$, and constrained the source of the
emission to radial distances within 2\,AU. \citet{Schindhelm2012} reported similar conditions for
the UV fluorescent \hm\ emission from a larger sample of T Tauri stars ($T(\hm)=2500\pm 1000$\,K,
$\log N(\hm)=19 \pm 1$\psqcm). The \hm\ line profiles reported in \citet{France2012} for single,
normal (non-transition) TTS suggest that the emission arises typically from radii $\sim$0.1--1\,AU.

\begin{figure}[t]\begin{center}
\includegraphics[width=3.25in]{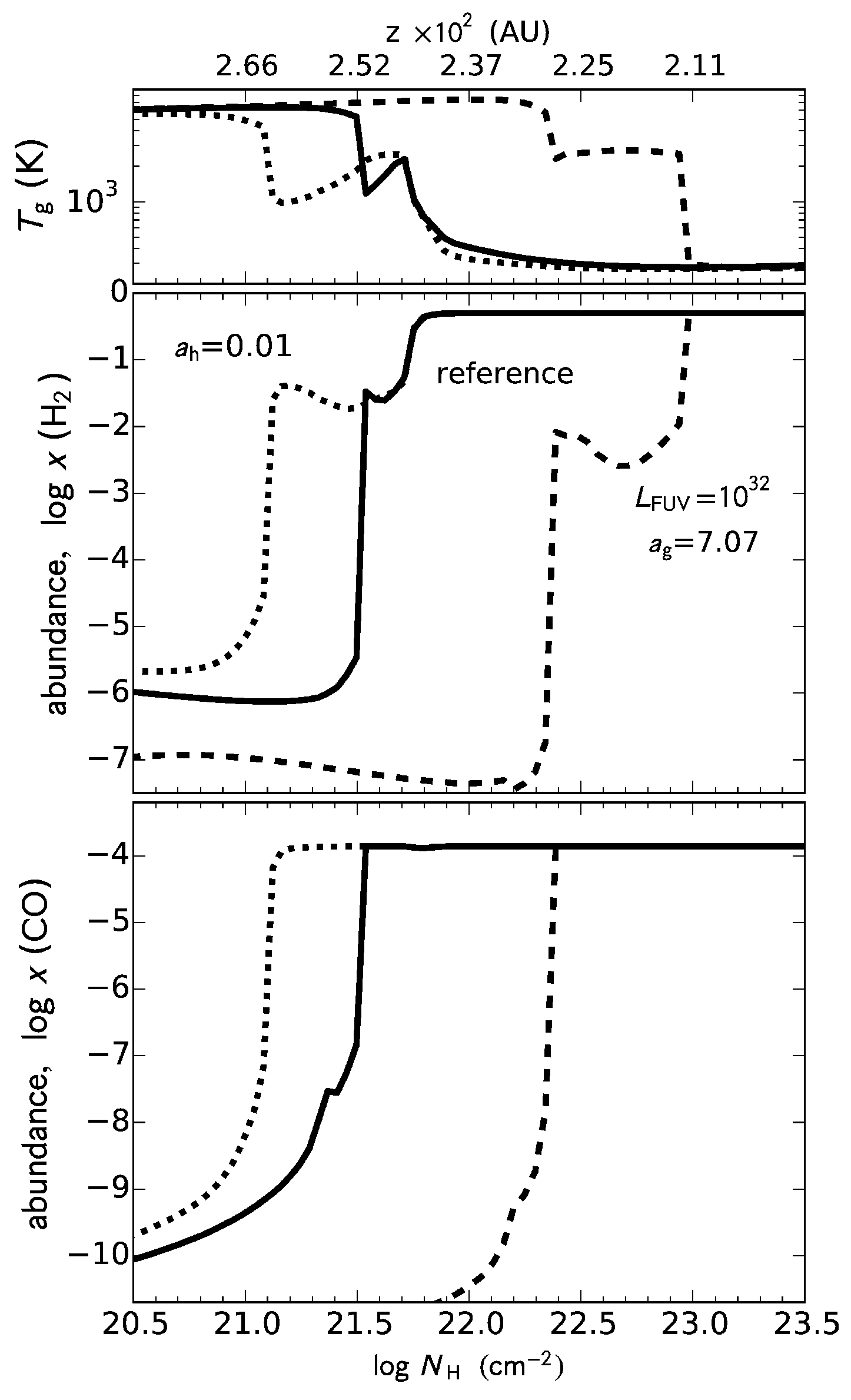}
\caption{\label{f:par_var}Vertical profiles of \Tg, $x(\hm)$, and $x({\rm CO})$ with different
values for \ah, \ag, and \LFUV\ at $r=0.24$\,AU. Profiles from a model with reduced mechanical
heating (dotted curves) have a transition higher in the atmosphere, while models with larger grains
and increased radiation (dashed curves) transition to molecular conditions deeper into the
atmosphere. A physical distance scale for the altitude above the midplane is given on the top axis.}
\end{center} \end{figure}

In the above studies, the T Tauri stars that show fluorescent \hm\ emission include sources that
have experienced significant grain settling relative to ISM values. The Taurus sources in Schindhelm
et al.\ (2012) span a range of 10\um\ silicate emission equivalent widths (0.22 -- 1.0) and MIR
colors \citep[$n_{13-25} = -0.4$ to 0.2;][]{Furlan2006}, which are consistent with grain area
reduction by a factor $\sim$100. The best fits in the \citet{Herczeg2004} modeling of the
fluorescent \hm\ emission from TW~Hya have the \hm\ mixed with little dust compared to an ISM grain
abundance (they assumed no dust). These conditions are similar to the reduced grain area adopted in
our models.

The hot \hm\ occurs in a region that is primarily atomic, $x(\hm)/x(\h)<$0.1, and mixed with a high
abundance of water, $x(\water)\sim 10^{-5}$. Thus, both scattering and photochemical heating are
important in this region of the atmosphere. Interestingly, \Lya\ reconstruction studies typically
assume that the UV fluorescent \hm\ occurs in a region with little \hi\ (Herczeg et al.; Schindhelm
et al.). Our results suggest that \hi\ scattering could play a larger role, potentially increasing
the likelihood of \Lya\ excitation of hot \hm, similar to the increased probability of
\Lya\ absorption by dust or water when \hi\ scattering occurs.

As noted in \citet{France2012}, the UV fluorescent \hm\ emission profiles indicate that the emission
arises from similar disk radii as the CO rovibrational emission from disks
\citep[e.g.,]{Salyk2011}. Our model also produces an enhanced column density of hot
($T>900$\,K) CO, with columns in the $10^{17} - 10^{18}\psqcm$ range, generally from within 1\,AU.
For models with large grain sizes (\ag=7.07\um) in the upper range of FUV luminosity
(\LFUV=0.025\LSun), a significant fraction of radiation penetrates deep into the disk and we see
columns of up to $2\times 10^{18} - 8\times 10^{18}\psqcm$ of hot CO within 0.5\,AU
(Figure~\ref{f:par_var}). These values compare favorably to the CO rovibrational emission properties
of T~Tauri disks \citep[e.g.,][]{Najita2003}. \citet{Salyk2011} reported line-of-sight CO emission
columns of $10^{18} - 10^{19}\psqcm$ and excitation temperatures of 900--1600\,K for a sample of Taurus
disks; the small CO emitting areas and the width of the emission lines are consistent with much of
the emission arising from $<0.4$\,AU. Thus, the enhanced hot \hm\ and CO that we find in the hot
atomic and warm irradiated layers of our model may help to explain the UV fluorescent \hm\ and CO
rovibrational emission that is commonly observed from T Tauri disks.

The study by \citet{Nomura2007} provides an interesting counterpoint to the current study.   In
their pioneering work on the excitation of \hm\ in disks, \citet{Nomura2007} illustrated how
protoplanetary disks that are irradiated by stellar UV and X-rays could produce detectable \hm\
emission at UV and IR wavelengths. There are many differences between the two models, e.g.,  the
chemical model used here is more detailed and  molecular shielding is included, but hydrostatic
equilibrium is not enforced and the \hm\ level populations are not calculated. However, one of the
most interesting differences between the two models is the way in which the stellar UV heats the
gas.  In \citet{Nomura2007}, the gas is heated by UV photons through grain photoelectric heating.
As grains settle out of the disk atmosphere, grain photoelectric heating is reduced, the temperature
of the gaseous atmosphere drops, and the \hm\ emission declines in strength. These results led
Nomura et al.\ to conclude that \hm\ emission will be strongest from disks with an abundant small
grain population, i.e., disks that have experienced little grain growth and settling. In contrast,
\hm\ emission is detected commonly from TTS, the majority of which have experienced significant
grain settling.

In our model, UV photons heat the gas primarily through photodissociation of OH and
\water. While photoelectric heating is included in our calculation, it plays a limited role
because of the reduced abundance of small grains in the atmosphere. In this way, heating by UV
remains strong even as grains settle out of the atmosphere. Indeed, as the grains settle, the FUV
radiation penetrates deeper, and larger columns of hot \hm\ and CO are produced.

Our models complement the disk model of \citet{Du2014} in both the methods employed and the
astrophysical issues that are addressed. In \citet{Du2014}, \Lya\ propagation is treated in a much
more sophisticated way and photochemical heating is included in a simplified way compared to the
methods used here. As in AGN14, \citet{Du2014} focussed on the distribution of warm water and
primarily on the region $> 1$\,AU in contrast to the smaller radii considered here. While
\citet{Du2014} did not address in their work the origin of fluorescent \hm, the methods used in
their work could be used to investigate this question in greater detail.

In summary, we find that the \Lya\ component of the FUV radiation field of young stars has a
significant impact on the thermal structure of disk atmospheres. \Lya\ photons scattered by \hi\ at
the top of the disk atmosphere deposit their energy deeper in the disk atmosphere than radially
propagating FUV continuum photons. In addition, scattering by \hi\ throughout the upper disk
atmosphere can cause much of the \Lya\ energy to be deposited over a restricted range in column
density, which leads to a surface layer of hot \hm. The temperature in the layer is high enough
($\sim$2000\,K) to thermally excite the \hm\ to vibrational levels from which they can be fluoresced
by \Lya\ to produce the UV fluorescent \hm\ emission that is characteristic of accreting young
stars. \Lya\ irradiation also leads to a layer of warm CO ($>$900K) in the inner disk atmosphere
that has a column density similar to that inferred for 4.7\um\ rovibrational CO emission from young
stars. The high \hm\ and CO temperatures are primarily the result of photochemical heating (by
\water, OH and \hm\ formation), a process that offers an efficient way to tap energy of the stellar
UV field when grains have settled out of disk atmospheres and photoelectric heating is diminished.

We have investigated the impact of \Lya\ heating on a particular disk atmosphere model. Given the
significant impact of \Lya\ irradiation on the thermal structure of the disk atmosphere, it is
important to examine its role relative to other processes over a range of disk radii.  More
generally, we have investigated the impact of \Lya\ heating in the context of one particular disk
atmosphere model. Further studies of the impact of \Lya\ heating under a wider range of disk
conditions would be useful to understand the general nature of this process.

\section*{Acknowledgments}

We acknowledge helpful conversations with Greg Herczeg and Kevin France, and support from the
following NASA grants: NNX15AE24G (Exoplanets Research Program), NNG06GF88G (Origins), 1367693
({\it Herschel} DIGIT), and Agreement No. NNX15AD94G for the program ``Earths in Other Solar
Systems''. This work was performed in part at the Aspen Center for Physics, which is supported by
National Science Foundation grant PHY-1066293.

\appendix

\section{FUV photochemical heating of O$_2$}

The photochemical heating discussed in Sec.~2 is based on GN15. However, the examples treated there
do not include $\otwo$ which can play a significant role in the FUV heating of the molecular layer.
The photochemical heating defined in Eq.~1 is composed of a direct and a chemical part, 
\be Q_{\rm
phchem}= Q_{\rm dir} + Q_{\rm chem}, 
\ee 
where $Q_{\rm dir}$ comes from the kinetic energy of the
dissociation fragments (two O atoms in this case), and $Q_{\rm chem}$ from their chemical energies.
As emphasized by GN15, photochemical heating is dependent on the density because some of it arises
from  excitation of the products and only leads to heating if the density is high enough for
collisional de-excitation of the excited levels to occur. In the present application to the
molecular layer, the densities are high enough ($> 10^{10} \pcc$) that it is a good approximation to
assume that essentially all excitation goes into heating.

The UV absorption cross section of $\otwo$ is well known. It is dominated by the Schumann-Runge
continuum from 1300-1800\,\AA, well measured by \citet{Yoshino2005}. Over much of this band, one of
the atoms is produced in the $^1$D$_2$ level \citep{Lee2000,Lee2001}. Following GN15, we assume that
all of the excitation energy, $E(^1{\rm D}_2) = 1.98$\,eV, is available for heating. Below
1400\,\AA, the branching ratio $b$ for the $^1$D$_2$ level is less than one, but this only reduces
the mean heating from its collisional de-excitation by $\sim 10\%$. The $^1$S$_0$ level at 4.19\,eV
is also produced below 1100\,\AA, but it quickly decays to the $^1$D$_2$ level. In any case, these
wavelengths are heavily blocked by $\hm$ and CO self-shielding, and we ignore $\otwo$ heating in the
900-1100\,\AA\ band.

The direct heating has been obtained by calculating the the mean value of the quantity,
\be
\Delta E_{\rm dir} = h\nu - D(\otwo) - b \, E(^1{\rm D}_2),
\ee 
in each of ten 100\,\AA\ bands from 1100-2400\,\AA; the dissociation energy of 
$\otwo$ has been set to $D(\otwo) = 5.12$\,eV, and the mean value of $b$ is 
$\bar{b} = 0.90$. The result is $Q_{\rm dir} = 1.6$\,eV. 
The chemical energy is obtained from the reactions, initiated by each O atom,
\be
\label{R:O+H2}
{\rm O} + \hm \, \ra \, \oh + \h,				
\ee
and,
\be
\label{R:OH+H2}
\oh + \hm \, \ra \, \water + \h,			 
\ee
which are are equivalent to,
\be
\label{Oheateq}
{\rm O} + 2\hm \, \ra \, \water + 2\h, 
\ee 
with a net energy yield of 0.57\,eV. Thus the net chemical energy is $Q_{\rm chem} = 2(0.57 +
\bar{b}\,1.98)$\,eV or 4.7\,eV, and the total photochemcial heating is $Q_{\rm phchem}(\otwo) =
6.3$\,eV. For the same conditions, the photochemical heating of \water\ and OH are 2.1\,eV and
5.5\,eV, respectively. The heating rates in Figure 1 reflect these values together with the
abundances of the oxygen molecules.



\setcounter{table}{0}
\renewcommand{\thetable}{A\arabic{table}}
\begin{deluxetable}{l*{4}c}
\tablecaption{\label{t:phchem}Photochemical heating energies and cross sections\tablenotemark{a}}
\tablehead{ X\tablenotemark{b} &  
$Q_{\rm cont}$ & 
$Q_{\Lya}$   & 
$\sigma_{\rm X}(\Lya)$\tablenotemark{c} & 
Refs.\tablenotemark{d}  
}
\startdata 
H$_2$O     & 2.1     & 1.6     &  12      &  1  \\
OH         & 5.5     & 6.4     &  1.8     &  2  \\
O$_2$      & 6.3     & \nodata &  0.01    &  3  \\ 
H$_2$      & 12.5    & \nodata &  \nodata &  4  \\
CO         & 8.7     & \nodata &  \nodata &  5  \\
C          & 8.0     & \nodata &  \nodata &  6  \\ 
N$_2$      & \nodata & \nodata &  \nodata &  7  
\enddata
\tablenotetext{a}{Units for photochemical heating energies are eV per photodissociation 
and cross sections are in $10^{-18}\, {\rm cm}^2$.}
\tablenotetext{b}{The species that contribute to FUV continuum photochemical heating have numerical
values in second column, $Q_{\rm FUV}$, and those considered for \Lya\ photochemcial heating
have values listed in the third column for $Q_{\Lya}$.}
\tablenotetext{c}{\Lya\ cross sections from \citet{vanDishoeck2006}. 
                  The molecules than cannot be dissociated from their ground state by 
                  \Lya\ do not have cross sections listed.}
\tablenotetext{d}{References for FUV continuum cross sections:
(1) \citet{Lee1986, Parkinson2003, Mota2005};
(2) \citet{vDD84};
(3) \citet{Yoshino1992, Yoshino2005}; 
References for line shielding:
(4) \citet{DB96}; 
(5) \citet{Visser2009};
(6) \citet{McGuire1968};
(7) \citet{Li2013}.
} 
\end{deluxetable}

\end{document}